# An Exact Method using Quantum Theory to Calculate the Noise Figure in a Low Noise Amplifier

Ahmad Salmanogli
Çankaya University, Engineering faculty, Electrical and Electronic Department, Ankara, Turkey

*Abstract*— In this article, a low noise amplifier is quantum mechanically analyzed to study the behavior of the noise figure. The analysis view is changed from the classic to quantum, because using quantum theory produces some degrees of freedom, which may be ignored when a circuit is analyzed using a classical theory. For this reason, the associated Lagrangian is initially derived for the circuit and then using Legendre transformation and canonical quantization procedure the classical and quantum Hamiltonian are derived, respectively. Consequently, the dynamic equation of motion of the circuit is introduced by which all of the circuit measurable observations such as voltage and current fluctuations are calculated. As an interesting point of this study, the low noise amplifier is deliberately supposed as two oscillators connecting to each other sharing the mutual specifications and accordingly the voltage and current are expressed in terms of the oscillator's photon number. As a result, one can analyze the critical quantity such as the noise figure in terms of the oscillator's photon number and also the photons coupling between oscillators. The latter mentioning term is considered as a factor to engineering the amplifier critical quantities. Additionally, the considered circuit is designed and classically simulated to testify the derived results using the quantum theory.

## I. INTRODUCTION

One of the critical parts of a Radar system [1],[2] contributes to the receiver subsystem as a very weak echo gets back from a higher altitude. The detection of such signals is very difficult, because the detecting signal power level equals the noise level [1],[2]. To enhance the radar receiver sensitivity, the related noise power of this subsystem should be limited as much as possible. For this subsystem, a low noise amplifier (LNA) [3]-[6] has been considered as an indispensable part, which can effectively amplify a very weak receiving signal, and also add a noise with a minimum level to the signal. In fact, in a radar receiver this is LNA that should dominate the sensitivity. However, the design of a LNA is mostly challenging. The LNA design contains some unusual trade-offs among noise figure (NF), linearity, gain, impedance matching, and also power consumption [3]-[6]. However, the designer mainly focuses on LNA related NF, which is a very critical factor in the detection of a very weak signal. This is due to the fact that the noise performance of a LNA can easily affect the receiver performance as well. There are some different types of LNA circuit such as common source LNA (CS-LNA) [7]-[9] and common gate LNA (CG-LNA) [7], [10]-[11]. The former one has been widely used due to its very good noise performance, while the latter one achieves a very good impedance matching and in contrast has a poor NF. For instance, to improve CG-LNA noise performance, the capacitive cross-coupling technique has been employed [3],[7]. Also, some additional inductors are added at the main transistor's drain to improve the noise performance by cancelling out the parasitic capacitance effect [5]-[7]. Additionally, the capacitance connected between gate and drain of the transistor in a cascade LNA [7],[12], which is mainly used to improve the LNA linearity [5]-[8] can help to minimize NF. So far, some LNA design circuits are cited in which the aim is to improve the performance of the LNA. However, there are so many articles (not cited in this study) that have investigated LNA performance enhancing due to its priority in systems such as radar. Beside this, there are some other applications such as quantum radars [13]-[15] that need to use an Ultra-low noise amplifier. This means that NF is very important in such an application. That is why we specifically focus on this factor in this study and try to analyze this quantity from the other and exact point of view called quantum theory [16]-[17]. In contrast to the classical point of view that the LNA has been analyzed so far, this study investigates LNA using quantum theory. It is supposed that quantum theory gives some extra degrees of freedom that the designed circuit becomes fully understandable. This study starts with an important assumption that the designed LNA operates like two coupled oscillators to each other. It means that all of the LNA important properties such as NF can be affected through coupling the mentioned oscillators to each other. Thus, this is the oscillator's photon number that affects the system performance. To completely understand the LNA operating, the circuit degrees of freedom such as voltage and current are derived and expressed in terms of the oscillator's photon numbers. Additionally, using the current and voltage definitions, other parameters such as power gain and power consumption can be defined in the same way.

## II. THEORY and BACKFROUND

A typical LNA (CS with a capacitive feedback) is schematically illustrated in Fig. 1a. In this circuit $V_{rf}$ as an input signal operating at RF frequency excites the circuit biased with $V_d$. In this circuit, a MOSFET transistor indicated with label "Q" is a non-linear element. In addition, the LNA equivalent circuit at RF frequency is schematically shown in Fig. 1b. In this circuit, the capacitors arising at high frequencies such as $C_{gs}$ and $C_{gd}$ are regarded and also non-linear elements indicating with $i_{ds}$ are defined as a dependent current source controlled by voltage $V_{gs}$ (dropping voltage across $C_{gs}$). The current can be expressed in terms of $V_{gs}$ as $i_{ds} = V_{gs}\partial i_{ds}/\partial V_{gs} + V_{gs}^2 \partial^2 i_{ds}/\partial V_{gs}^2 + V_{gs}^3 \partial^3 i_{ds}/\partial V_{gs}^3 = g_m V_{gs} + g_{m2} V_{gs}^2 + g_{m3} V_{gs}^3$ [5],[6], where $g_m$ is a linear term standing for the intrinsic transconductance of the transistor and $g_{m2}$, $g_{m3}$ are the non-linear quantities used to approximately model the transistor as a non-linear element. Additionally, in the equivalent circuit, the current sources defined as $\bar{I}_s = \bar{I}_{s0} + \sqrt{\bar{I}_R^2}$ and $\bar{I}_d = \bar{I}_{d0} + \sqrt{\bar{I}_d^2}$, where $\bar{I}_{s0}$ and $\bar{I}_{d0}$ indicate DC bias current and $\bar{I}_R^2 = 4KTR_s$ stands for the input-induced noise [5-9]. That is because of any resistors appearing in the LNA. In this equation, K and T are the Boltzmann's constant and operating temperature, respectively. Finally, $\bar{I}_d^2 = 4KT\gamma g_m$ is the thermal noise [3]-[5], where $\gamma$ is the empirical constant with a typical value of $\gamma = 2/3$ for a long channel MOSFET transistors.

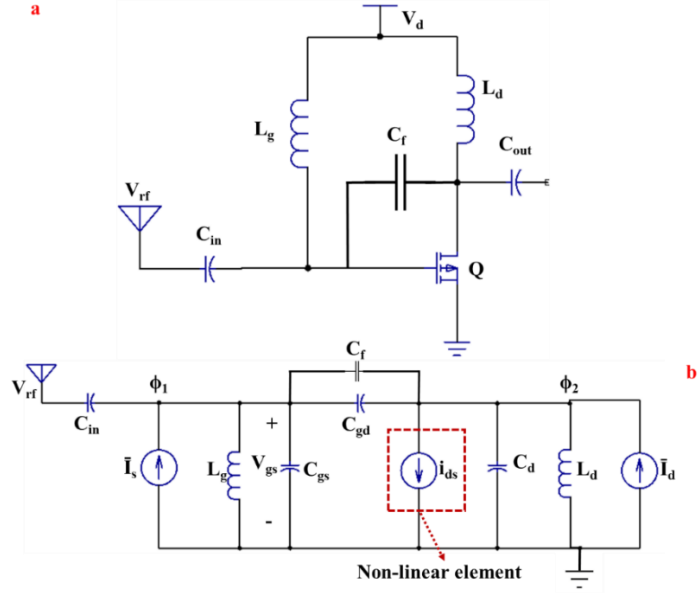

Fig. 1. Schematic of (a) simple LNA and (b) the contributed equivalent circuit at RF frequencies; in the equivalent circuit $i_{ds}$ stands for the non-linear element and $\bar{I}_s$ and $\bar{I}_d$ contain the circuit DC bias and noise effects.

For analysis of the circuit with the full quantum theory, the essential nodes fluxes are defined as the coordinates (input and output node) and also loop charge as the contributed momentum conjugate variable. The node fluxes ($\varphi_1$, $\varphi_2$) and loop charges ($Q_1$,$Q_2$) are related to voltage and current. Circuit analysis begins with the definition of the Lagrangian [16]-[17] (difference between the electric and magnetic energy stored in the elements) as:

$$L_c = \frac{C_{gs}}{2}\dot{\varphi}_1^2 - \frac{1}{2L_g}\varphi_1^2 + \frac{C_d}{2}\dot{\varphi}_2^2 - \frac{1}{2L_d}\varphi_2^2 + \frac{C_{gd}}{2}(\dot{\varphi}_1 - \dot{\varphi}_2)^2$$
$$+ \dot{\varphi}_2(\bar{I}_d + g_m\dot{\varphi}_1 + g_{m2}\dot{\varphi}_1^2 + g_{m3}\dot{\varphi}_1^3) + \varphi_1\bar{I}_s + \frac{C_{in}}{2}(\dot{\varphi}_1 - V_{rf})^2$$

(1)

where $V_1 = d\varphi_1/dt$ and $V_2 = d\varphi_2/dt$; from the circuit it is clear that $V_1 = V_{gs}$. The associated classical Hamiltonian can be obtained using Legendre transformation $H(\varphi_k, Q_k) = \sum_k (\dot{\varphi}_k \cdot Q_k) - L_c$, where $Q_k$ are the conjugate variables of the coordinate $\varphi_k$ satisfying the Poisson bracket $\{\varphi_k, Q_{k'}\} = \delta_{kk'}$ and calculated as $Q_k = \partial L_c / \partial(\partial \varphi_k/\partial t)$. The classical Hamiltonian using the Legendre transformation is presented as:

$$H_c = \frac{(C_{in} + C_{gs} + C_{gd})}{2}\dot{\varphi}_1^2 + \frac{1}{2L_g}\varphi_1^2 + \frac{(C_d + C_{gd})}{2}\dot{\varphi}_2^2 + \frac{1}{2L_d}\varphi_2^2$$
$$- C_{gd}\dot{\varphi}_1\dot{\varphi}_2 + g_{m2}\dot{\varphi}_1^2\dot{\varphi}_2 + 2g_{m3}\dot{\varphi}_1^3\dot{\varphi}_2 - \bar{I}_s\varphi_1 - \bar{I}_d\varphi_2 - \frac{C_{in}}{2}V_{rf}^2$$

(2)

In this equation, the term $2g_{m3}\varphi_2 \partial^3\varphi_1/\partial t^3$ is approximated with $2g_{m3L}\varphi_2 \partial^2\varphi_1/\partial t^2$ to simplify the Hamiltonian. The approximation term is given by $g_{m3L} = g_{m3}\varphi_{1dc}$, where $\varphi_{1dc} = \partial\varphi_1/\partial t]^{DC}$ is the first derivative of $\varphi_1$ at the approximation point. The final step is to apply the canonical conjugate quantization procedure on the classical degrees of freedom $[\varphi_k, Q_k] = i\hbar$, yielding the quantum Hamiltonian. For the sake of simplicity and a clear analysis, the Hamiltonian is divided into two linear and non-linear parts. The linear part is given by:

$$H_L = \frac{1}{2C_{q1}}Q_1^2 + \left(\frac{1}{2L_g} + \frac{g_m^2}{2C_{p1}}\right)\varphi_1^2 + \frac{1}{2C_{q2}}Q_2^2 + \left(\frac{1}{2L_d} + \frac{g_m^2}{2C_{p2}}\right)\varphi_2^2 + \frac{g_m^2}{2C_{p1p2}}\varphi_1\varphi_2 + \frac{P_1 g_m V_{rf}}{2}\varphi_1 + \frac{1}{2C_{q1q2}}Q_1 Q_2$$
$$+ \frac{g_m}{2C_{q1p1}}Q_1\varphi_1 + \frac{g_m}{2C_{q2p2}}Q_2\varphi_2 + \frac{g_m}{2C_{q1p2}}Q_1\varphi_2 + \frac{g_m}{2C_{q2p1}}Q_2\varphi_1 + \frac{P_2 g_m V_{rf}}{2}\varphi_2 + \frac{G_1 V_{rf}}{2}Q_1 + \frac{G_2 V_{rf}}{2}Q_2$$

(3)

where $C_{q1}$, $C_{q2}$, $C_{p1}$, $C_{p2}$, $C_{q1q2}$, $C_{q1p1}$, $C_{q1p2}$, $C_{q2p1}$, $C_{q2p2}$, $P_1$, $P_2$, $G_1$ and $G_2$ are constants defined in Appendix A. In this equation, we used two definitions as the following $1/2L_{g'} \equiv (1/2L_g + g_m^2/2C_{p1})$ and $1/2L_{d'} \equiv (1/2L_d + g_m^2/2C_{p2})$; at each one the second term indicates the effect of the intrinsic transconductance and also coupling capacitors on both gate and drain connecting inductors. In other words, the inductors connected to the transistor can be manipulated due to the coupling effect. The important point here is that one can easily find from the Hamiltonian expressed in Eq. 3 that the equation contains two oscillators; the first oscillator is connected to the gate and the other is connected to the drain of the transistor and oscillate respectively with $\omega_1 = 1/\sqrt{(L_{g'} \cdot C_{q1})}$ and $\omega_2 = 1/\sqrt{(L_{d'} \cdot C_{q2})}$. Other terms in Eq. 3 such as $Q_1Q_2$, $Q_1\varphi_1$, $Q_2\varphi_1$, $Q_1\varphi_2$ show the coupling between oscillators, meaning that the oscillators share the energy between themselves. The last terms as $Q_1$, $\varphi_1$, $Q_2$, $\varphi_2$ declare the coupling of the RF source to the contributed oscillators. Beside of the linear Hamiltonian expressed in Eq. 3, the non-linear Hamiltonian is presented as:

$$H_{NL} = (g_{m2} + 2g_{m3L})\begin{bmatrix}\left\{\frac{V_{rf}}{2C'_{q1p2}}Q_1\varphi_2 + \frac{V_{rf}}{2C'_{q2p2}}Q_2\varphi_2 + \frac{g_m V_{rf}}{2C'_{p1p2}}\varphi_1\varphi_2 + C_{11}^2 C_{in}^2 V_{rf}^2 \varphi_2\right\}_{NL2} \\ + \left\{\begin{array}{l}C_{11}^2 Q_1^2\varphi_2 + C_{12}^2 Q_2^2\varphi_2 + 2C_{11}C_{12}Q_1 Q_2\varphi_2 + C_{11}^2 g_m^2 \varphi_1^2\varphi_2 \\ + 2C_{11}^2 g_m Q_1\varphi_1\varphi_2 + 2C_{11}C_{12}g_m Q_2\varphi_1\varphi_2\end{array}\right\}_{NL3}\end{bmatrix}$$

(4)

where $C'_{q1p2}$, $C'_{p1p2}$, $C'_{q2p2}$, $C_{11}$, $C_{12}$, and $C_{22}$ are constants and defined in Appendix A. One can consider $H_{NL}$ in Eq. 4 as $H_{NL} = H_{NL2} + H_{NL3}$, where $H_{NL3}$ regarded as the perturbation term and its effect on the energy eigenvalue and eigenstates will be studied. In the following, the total Hamiltonian is considered as $H = H_0 + H_p$, where $H_0 = H_L + H_{NL2}$ stands for un-perturbed Hamiltonian and $H_p = H_{NL3}$ indicates the perturbed Hamiltonian. To derive the dynamic equation of motion of the circuit to study the parameters such as voltage and current from quantum theory point of view, it needs to define the Hamiltonian in terms of the creation and annihilation operators. For this reason, the coordinate parameters ($\varphi_1$, $\varphi_2$) and the related momentum conjugate ($Q_1$, $Q_2$) are expressed in terms of the ladder operators. One can easily re-express the Hamiltonian in terms of the ladder operators using the quantization procedure $Q_1 = -i(a_1 - a_1^+) \cdot \sqrt{(\hbar/2Z_1)}$, $\varphi_1 = (a_1 + a_1^+) \cdot \sqrt{(\hbar Z_1/2)}$ and $Q_2 = -i(a_2 - a_2^+)\sqrt{(\hbar/2Z_2)}$, $\varphi_2 = (a_2 + a_2^+)\sqrt{(\hbar Z_2/2)}$, where $(a_k, a_k^+)$ $k = 1, 2$ are the first and second oscillator's ladder operators. Also, the contributed impedance for each oscillator are expressed as $Z_1 = \sqrt{(L_{g'}/C_{q1})}$ and $Z_2 = \sqrt{(L_{d'}/C_{q2})}$. Thus, $H_0$ is introduced as expressed in Eq. 5:

$$H_0 = \left\{\hbar\omega_1\left(a_1^+ a_1 + \frac{1}{2}\right) + \hbar\omega_2\left(a_2^+ a_2 + \frac{1}{2}\right) - \frac{\hbar}{4C_{q1q2}\sqrt{Z_1 Z_2}}(a_1 - a_1^+)(a_2 - a_2^+)\right.$$
$$- \frac{i\hbar \cdot g_m}{4C_{q1p1}}(a_1 - a_1^+)(a_1 + a_1^+) - \frac{i\hbar \cdot g_m}{4C_{q2p2}}(a_2 - a_2^+)(a_2 + a_2^+) - \frac{i\hbar \cdot g_m}{4C_{q1p2}}\sqrt{\frac{Z_2}{Z_1}}(a_1 - a_1^+)(a_2 + a_2^+)$$
$$+ \frac{\hbar \cdot g_m^2}{4C_{q1p2}}\sqrt{Z_2 Z_1}(a_1 + a_1^+)(a_2 + a_2^+) - \frac{i\hbar \cdot g_m}{4C_{q2p1}}(a_1 + a_1^+)(a_2 - a_2^+) + \frac{P_1 V_{rf} g_m}{2}\sqrt{\frac{\hbar Z_1}{2}}(a_1 + a_1^+)$$
$$\left.+ \frac{P_2 V_{rf} g_m}{2}\sqrt{\frac{\hbar Z_2}{2}}(a_2 + a_2^+) - \frac{iG_1 V_{rf} g_m}{2}\sqrt{\frac{\hbar}{2Z_1}}(a_1 - a_1^+) - \frac{iG_2 V_{rf} g_m}{2}\sqrt{\frac{\hbar}{2Z_2}}(a_2 - a_2^+)\right\}_L$$
$$+ (g_{m2} + 2g_{m3L}) \times \left\{\frac{\hbar g_m V_{rf}}{4C'_{p1p2}}\sqrt{Z_2 Z_1}(a_1 + a_1^+)(a_2 + a_2^+) + C_{11}^2 C_{in}^2 V_{rf}^2 \sqrt{\frac{\hbar Z_2}{2}}(a_2 + a_2^+)\right.$$
$$\left.- \frac{i\hbar g_m V_{rf}}{4C'_{q1p2}}\sqrt{\frac{Z_2}{Z_1}}(a_1 - a_1^+)(a_2 + a_2^+) - \frac{i\hbar V_{rf}}{4C'_{q2p2}}(a_2 - a_2^+)(a_2 + a_2^+)\right\}_{NL2}$$

(5)

where subscripts "L" and "NL2" stand for the linear and nonlinear parts, respectively. In the Hamiltonian expressed in Eq. 5, all of the constant terms are ignored to simplify the equation. Also, the perturbation Hamiltonian is given by:

$$H_p = (g_{m2} + 2g_{m3L}) \times \left\{\frac{-\hbar}{2Z_1}C_{11}^2(a_1 - a_1^+)^2 - \frac{\hbar}{2Z_2}C_{12}^2(a_2 - a_2^+)^2 + \frac{\hbar Z_1 g_m^2}{2}C_{11}^2(a_1 + a_1^+)^2\right.$$
$$- \frac{\hbar}{2\sqrt{Z_1 Z_2}}2C_{12}C_{11}(a_1 - a_1^+)(a_2 - a_2^+) - \frac{i\hbar}{2}(2C_{11}^2 g_m)(a_1 - a_1^+)(a_1 + a_1^+)$$
$$\left.- \frac{i\hbar}{2}(2C_{11}C_{12}g_m)\sqrt{\frac{Z_1}{Z_2}}(a_2 - a_2^+)(a_1 + a_1^+)\right\}\sqrt{\frac{\hbar Z_2}{2}}(a_2 + a_2^+)$$

(6)

Table 1. Data used to simulate the LNA using quantum theory

|  | Stands for |  |
|---|---|---|
| W | Transistor channel width | 300 um |
| $L_t$ | Transistor channel length | 50 um |
| $t_{SiO2}$ | $SiO_2$ thickness | 200 nm |
| $L_{ov}$ | Overlapping length | $0.1 \times L_t$ |
| $T_c$ | Operational temperature | 4 K |
| $\gamma$ | Empirical constant | 2/3 |
| $C_f$ | Feedback capacitor | 0.2 pF |
| $C_{in}$ | Input capacitance | 1.8 pF |
| $C_d$ | Drain capacitance | 0.08 pF |
| $L_g$ | Gate inductance | 1.2 nH |
| $L_d$ | Drain inductance | 0.95 nH |
| $g_{m2}$ | Second order non-linearity [5,6] | 250 mA/V$^2$ |
| $g_{m3}$ | Third order non-linearity [5,6] | 1300 mA/V$^3$ |

III. RESULTS and DISCUSSIONS

*A. The oscillators Energy dispersion*

In this section, we tried to analyze the energy associated with each oscillator and examine the effect of the non-linear part on the energy levels (using Table. 1 according to ATF54143 transistor's data). It supposes $|j_1\rangle$ and $|j_2\rangle$ as the energy state of the first and second oscillator, respectively. Therefore, using Hamiltonian expressed in Eq. 5, one can calculate the energy of the oscillators using $E_j = \langle j_i|H_0|j_i\rangle$, where $j_i$ = 1,2. Thus, the oscillators associated energies due to the un-perturbed Hamiltonian ($H_0$) are given by:

$$E_{j_1} = \langle j_1|H_0|j_1\rangle = \hbar\omega_1\left(j_1+\frac{1}{2}\right) - \frac{i\hbar}{2}\cdot\frac{g_m}{2C_{q1p1}}(j_1),$$

$$E_{j_2} = \langle j_2|H_0|j_2\rangle = \hbar\omega_2\left(j_2+\frac{1}{2}\right) - \frac{i\hbar}{2}\cdot\frac{g_m}{2C_{q2p2}}(j_2) - \frac{i\hbar}{2}\cdot\frac{V_{rf}}{2C'_{q2p2}}(j_2)$$

(7)

Eq. 6 clearly shows that each oscillator is affected by the associated energy (first term) and also by the intrinsic transconductance of the transistor (second term). It means that changing the gain of the current in the transistor should impact the energy levels of the oscillators. Also, $E_2$ is additionally influenced by the other factor that arose due the non-linearity induced in the transistor. This is a critical factor manipulated by the RF incident wave and can change the energy of the second oscillator. This factor will be discussed in detail in the following. Moreover, we considered the effect of the perturbed Hamiltonian on the energy levels of the oscillators and it found that $H_p = H_{NL3}$ doesn't affect the energy levels of the oscillators. It means $E_j^{(1)} = \langle j_i|H_p|j_i\rangle = 0$, where $E_j^{(1)}$ is the change in energy using the first order perturbation theory. The oscillator's energy dispersion is studied and simulated and the results are depicted in Fig. 2. In this figure, it is tried to depict the oscillators associated energy levels versus transistor intrinsic transconductance for different values of the RF incident wave. From Eq. 7, it is clear that $E_{j1}$ doesn't be affected by $V_{rf}$ amplitude, therefore, in Fig. 2a it just considers the effect of $j_1$. In contrast, Fig. 2b shows the effect of the $V_{rf}$ amplitude on $E_2$ energy. It is shown in Fig. 2b that increasing the incident wave amplitude ($V_{rf}$) manipulates $E_2$ but not $E_1$. This comes from the last term in $H_{0\{NL2\}}$ which controls the coupling of $V_{rf}$ to the second oscillator.

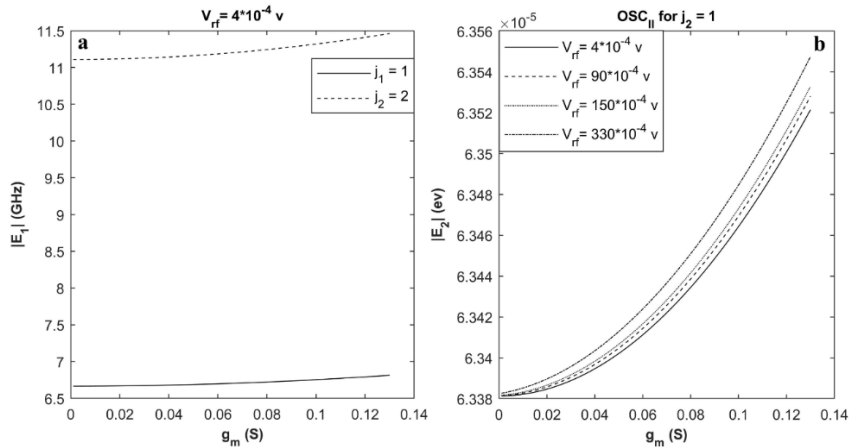

Fig. 2 Oscillators energy dispersion vs. $g_m$ (S), for different values of $V_{rf}$; a) absolute value of the energy of $OSC_I$ for different level of energy $j_1$ = 1 and 2, b) a) absolute value of the energy of $OSC_{II}$ for different value of $V_{rf}$ at $j_2$ = 1.

## B. Energy Eigen state due to the Perturbation Hamiltonian

Beside the energy (Eigen-values), the state (Eigenstate) of the oscillators can be dispersed. One can utilize the first perturbation theory to figure out about the oscillator's state changing due to the perturbation effect. The change of a typical state is calculated using first perturbation theory by: $|j>^{(1)} = \sum_{i \neq j} \{(<i|H_p|j>)/(E_i-E_j)\} \times |j>$, where $|j>$ is the pure state of the oscillators. The results of calculation are presented in Eq. 8 as:

$$|j_1\rangle^{(1)} = \sum_{i \neq j_1} \frac{\langle i|H_p|j_1\rangle}{E_i - E_{j1}} |j_1\rangle = 0$$
$$|j_2\rangle^{(1)} = \sum_{i \neq j_2} \frac{\langle i|H_p|j_2\rangle}{E_i - E_{j2}} |j_2\rangle = \frac{-\hbar}{2} C_{12}^2 \sqrt{\frac{\hbar}{2Z_2}} \left\{ \frac{|j_2-3\rangle}{E_{(j_2-3)} - E_{j_2}} + \frac{|j_2+3\rangle}{E_{(j_2+3)} - E_{j_2}} + \frac{(3j_2+1)|j_2-1\rangle}{E_{(j_2-1)} - E_{j_2}} + \frac{(2j_2+1)|j_2+1\rangle}{E_{(j_2+1)} - E_{j_2}} \right\}$$
(8)

This equation shows that $|j_1>$ doesn't be disturbed by $H_p$, while $|j_2>$ is strongly affected by the perturbation Hamiltonian. It is found that the state of the second oscillator $|j_2>$ is coupled to $|j_2 \pm 1>$ and $|j_2 \pm 3>$ due to the nonlinearity effect. For instance, if one fixes the second oscillator state at $|0>$ as a pure state, the final state of this oscillator is mixed to $|0>$ with $|1>$ and $|3>$; meaning that $H_p$ causes mixed states rather than a pure state. As an another example, for a pure state $|3>$, the final state that the oscillator can resonance in, is declared as $|0>$, $|2>$, $|4>$ and $|6>$; causes a mixing of state $|3>$ with states $|0>$, $|2>$, $|4>$ and $|6>$.

## C. Fluctuation of the voltage, current and transconductance

Using the Hamiltonian ($H_0$) expressed in Eq. 3 and Eq. 4, one can calculate the current and voltage as an important physical variable. The voltage and current operators ($V_1$ and $I_1$ for the first oscillator and $V_2$ and $I_2$ for the second oscillator) are calculated (using $V = [\phi, H_0]/i\hbar$, $I = [Q, H_0]/i\hbar$) and given by:

$$V_1 = \left\{ \frac{Q_1}{C_{q1}} + \frac{Q_2}{C_{q1q2}} + \frac{g_m\varphi_1}{2C_{q2p1}} + \frac{g_m\varphi_2}{2C_{q1p2}} + \frac{G_1 V_{rf}}{2} + \frac{V_{rf} g_N \varphi_2}{2C'_{q1p2}} \right\}$$
$$I_1 = -\left\{ \frac{\varphi_1}{L_{g'}} + \frac{g_m Q_1}{2C_{q1p1}} + \frac{g_m Q_2}{2C_{q2p1}} + \frac{g_m^2 \varphi_2}{2C_{p1p2}} + \frac{P_1 g_m V_{rf}}{2} + \frac{V_{rf} g_m g_N \varphi_2}{2C'_{p1p2}} \right\}$$
$$V_2 = \frac{Q_2}{C_{q2}} + \frac{Q_1}{C_{q1q2}} + \frac{g_m\varphi_1}{2C_{q2p1}} + \frac{g_m\varphi_2}{2C_{q2p2}} + \frac{G_2 V_{rf}}{2} + \frac{V_{rf} g_N \varphi_2}{2C'_{q2p2}}$$
$$I_2 = -\left\{ \frac{\varphi_2}{L_{d'}} + \frac{g_m Q_2}{2C_{q2p2}} + \frac{g_m Q_1}{2C_{q1p2}} + \frac{g_m^2 \varphi_1}{2C_{p1p2}} + \frac{P_2 g_m V_{rf}}{2} + \frac{V_{rf} g_m g_N \varphi_1}{2C'_{p1p2}} + \frac{V_{rf} g_N Q_1}{2C'_{q1p2}} + \frac{V_{rf} g_N Q_2}{2C'_{q2p2}} \right\}$$
(9)

Finally, the fluctuation of degrees of freedom (physical variables variance) are calculated using $\Delta V^2 = <V^2> - <V>^2$ and $\Delta I^2 = <I^2> - <I>^2$. The calculation results are expressed in Eq. 10 as:

$$\Delta V_1^2 = \frac{\hbar \omega_1}{2C_{Q1}}(2n_{1ph}+1) + \frac{\hbar \omega_2}{2C_{Q12}}(2n_{2ph}+1); \quad \Delta I_1^2 = \frac{\hbar \omega_1}{2L_{g1}}(2n_{1ph}+1) + \frac{\hbar \omega_2}{2L_{g12}}(2n_{2ph}+1)$$
$$\Delta V_2^2 = \frac{\hbar \omega_1}{2C_{Q21}}(2n_{1ph}+1) + \frac{\hbar \omega_2}{2C_{Q2}}(2n_{2ph}+1); \quad \Delta I_2^2 = \frac{\hbar \omega_1}{2L_{d21}}(2n_{1ph}+1) + \frac{\hbar \omega_2}{2L_{d2}}(2n_{2ph}+1)$$
(10)

where $n_{1ph}$ and $n_{2ph}$ are the first and second oscillators number of photons, respectively, and are calculated using Hamiltonian expressed in Eq. 5 by examining the expectation value of the $<a_k^+ a_k>$. The constants used in Eq. 10 including $C_{Q1}$, $C_{Q2}$, $C_{Q12}$, $C_{Q21}$, $L_{g1}$, $L_{g12}$, $L_{d1}$, and $L_{d21}$ are defined as:

$$\frac{1}{C_{Q1}} = \left\{ \frac{1}{C_{q1}} + \frac{g_m^2 Z_1^2 C_{q1}}{4C_{q1p1}^2} \right\}; \quad \frac{1}{C_{Q12}} = \left\{ \frac{C_{q2}}{4C_{q1q2}^2} + Z_2^2 C_{q2} \left( \frac{g_m}{2C_{q1p2}} + \frac{g_N V_{rf}}{2C'_{q1p2}} \right)^2 \right\}$$
$$\frac{1}{C_{Q2}} = \left\{ \frac{1}{C_{q2}} + Z_2^2 C_{q2} \left( \frac{g_m}{2C_{q2p2}} - \frac{g_N V_{rf}}{2C'_{q2p2}} \right)^2 \right\}; \quad \frac{1}{C_{Q21}} = \left\{ \frac{C_{q1}}{4C_{q1q2}^2} + \frac{g_m^2 Z_1^2 C_{q1}}{4C_{q2p1}^2} \right\}$$
(11)

and

$$\frac{1}{L_{g1}} = \left\{ \frac{1}{L_{g'}} + \frac{g_m^2 C_{q1}}{4C_{q1p1}^2} \right\}; \quad \frac{1}{L_{g12}} = \left\{ \frac{g_m^2 C_{q2}}{4C_{q2p1}^2} + Z_2^2 C_{q2} \left( \frac{g_m^2}{4C_{p1p2}} + \frac{2g_m g_N V_{rf}}{4C'_{p1p2}} \right)^2 \right\}$$
$$\frac{1}{L_{d2}} = \left\{ \frac{1}{L_{d'}} + C_{q2} \left( \frac{g_m}{2C_{q2p2}} - \frac{g_N V_{rf}}{2C'_{q2p2}} \right)^2 \right\}; \quad \frac{1}{L_{d21}} = \left\{ C_{q1} \left( \frac{g_m}{2C_{q1p2}} - \frac{g_N V_{rf}}{2C'_{q1p2}} \right)^2 + Z_1^2 C_{q1} \left( \frac{g_m}{2C_{p1p2}} - \frac{g_N V_{rf}}{2C'_{p1p2}} \right)^2 \right\}$$
(12)

where $g_{NL} = g_{m2} + 2g_{m3}L$. Eq. 10 clearly shows that the voltage and current fluctuation is strongly dependent on the number of photons of the oscillators. It means that this is the oscillator's number of photons to make the voltage and current fluctuations. The results expressed in Eq. 10 can be considered as the main impact of this work by which one can clearly observe the dependence of the fluctuation of the oscillators on the number of photons. This equation additionally shows that the photons created by an oscillator can couple to the other oscillator and change the specification of the oscillator. Accordingly, the coupling is done by a factor shown in Eq. 11 and Eq. 12. The latter mentioned equations demonstrate that the coupling factor is specifically manipulated by the transconductance of the transistor ($g_m$) and also the non-linear effect produced by the transistor ($g_N$). To get a better idea about the effect of the photons number on the voltage and current fluctuation, one can consider and study the oscillators' contributed number of photons plotting versus intrinsic transconductance and incident RF frequency ($\omega_{in}$). The results are illustrated in Fig. 3.

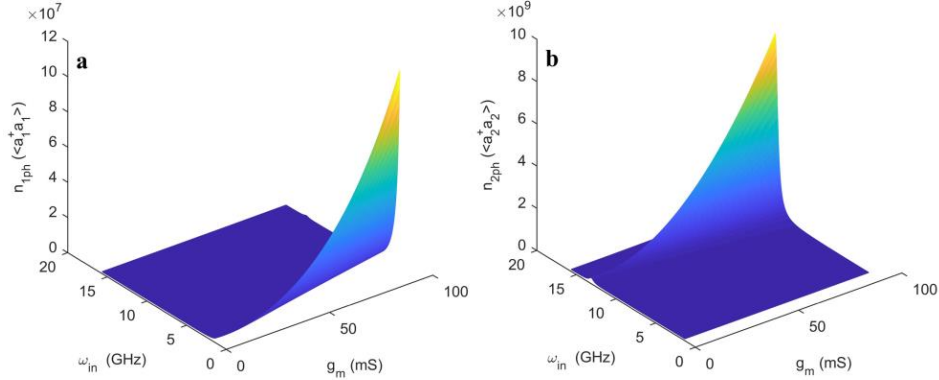

Fig. 3. Number of oscillators output photons vs. RF source angular frequency ($\omega_{in}$) and $g_m$, $V_{rf} = 3 \times 10^{-4}$ v, $C_f = 0.002$ pF a) $OSC_I$ and b) $OSC_{II}$

From the figures shown, it is clear to understand the effect of the design parameters on the oscillator's number of photons. In this simulation, the RF incident wave amplitude is assumed to be $V_{rf} = 4 \times 10^{-4}$ v. First of all, if one compares the amplitude of the number of the photons of the oscillators, it is clear that $n_{2ph} > n_{1ph}$; this is due to the fact that an LNA generally amplifies the input signals. Also, in each figure, increasing $g_m$ leads to photons increasing. This relates to the increase of the energy of oscillators. This point is illustrated in Fig. 2. Nonetheless, the significant point is $n_{2ph}$ amplification as $\omega_{in}$ is increased. This means that the OSC_II number of photons is directly affected by that factor.

Additionally, one can use Eq. 10 to derive the other important and critical factors in LNA like noise figure. Thus, using the general formula [4-8], NF is calculated in quantum area in terms of the oscillators photon number as:

$$NF = 1 + \frac{N_{device}}{G.N_{input}} = 1 + \frac{4KT\gamma g_m}{<\frac{\Delta I_2^2}{\Delta V_1^2}>.4KTR_s} = 1 + \frac{4\gamma g_m}{R_s} \left\{ \frac{\frac{\hbar\omega_1}{2C_{Q1}}(2n_{1ph}+1) + \frac{\hbar\omega_2}{2C_{Q12}}(2n_{2ph}+1)}{\frac{\hbar\omega_1}{2L_{d21}}(2n_{1ph}+1) + \frac{\hbar\omega_2}{2L_{d2}}(2n_{2ph}+1)} \right\} \quad (13)$$

where G is the power gain and $N_{device}$ and $N_{input}$ [3]-[9] are the device thermal noise and input noise, respectively. In Fig. 4, NF is simulated for the selected data as $V_{rf} = 4 \times 10^{-4}$ v and $C_f = 0.002$ pF. From Eq. 13, it is clear that everything relates to the oscillator's number of photons. It means that by controlling the number of photons one can deliberately manage NF for the designed LNA. From the results depicted in Fig. 3, we got the point that the crucial case is just the second oscillator, which is affected by $\omega_{in}$ and also $g_m$. Thus, the change of $\omega_{in}$ and $g_m$ leads to change $n_{2ph}$ as expressed in Eq. 13. As one can observe in Fig. 4, increasing $g_m$ causes an increase of NF. This is contributed to the power gain decreasing. However, the important point is the area on the figure indicated with a dashed-square. The mentioned odd behavior occurs where the incident frequency equals the summation of the two oscillators frequency $\omega_{in} = \omega_1 + \omega_2$. At that frequency NF is strongly decreased. This gives a clue to any designer when the main aim is to maintain the LNA related NF at the minimum level. This achievement is so important and in the following, we will show that this can just be fully predicted by quantum theory. To show the point, the LNA illustrated in Fig. 1 is simulated using an electronic based software to testify the result. The main aim is to compare the results attained using the quantum theory with the simulation result. For the LNA simulation, we used the data from Table. 1. The LNA PCB layout is depicted in Fig. 5a. If one compares Fig. 5a with Fig. 1a, it is clear that the inductors are a passive element in Fig. 1a is replaced with the Microstrip transmission line as a RF element. In this modeling, ATF54143 transistor is used, which is schematically illustrated as an inset figure in Fig. 5a. Fig. 5b shows the circuit transconductance versus incident wave frequency. This is a typical value for the circuit transconductance and it can be controlled and engineered for the desired amplitude on operational frequencies. However, it is not the aim of this study.

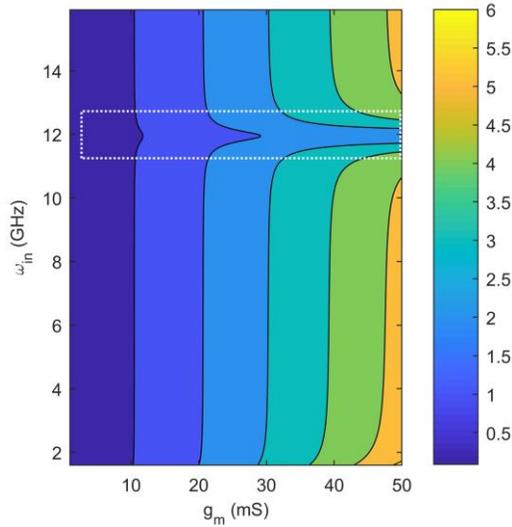

Fig. 4. NF vs. RF incident wave angular frequency (GHz) and intrinsic transconductance (S), $V_{rf} = 3 \times 10^{-4}$ v, $C_f = 0.08$ pF effect.

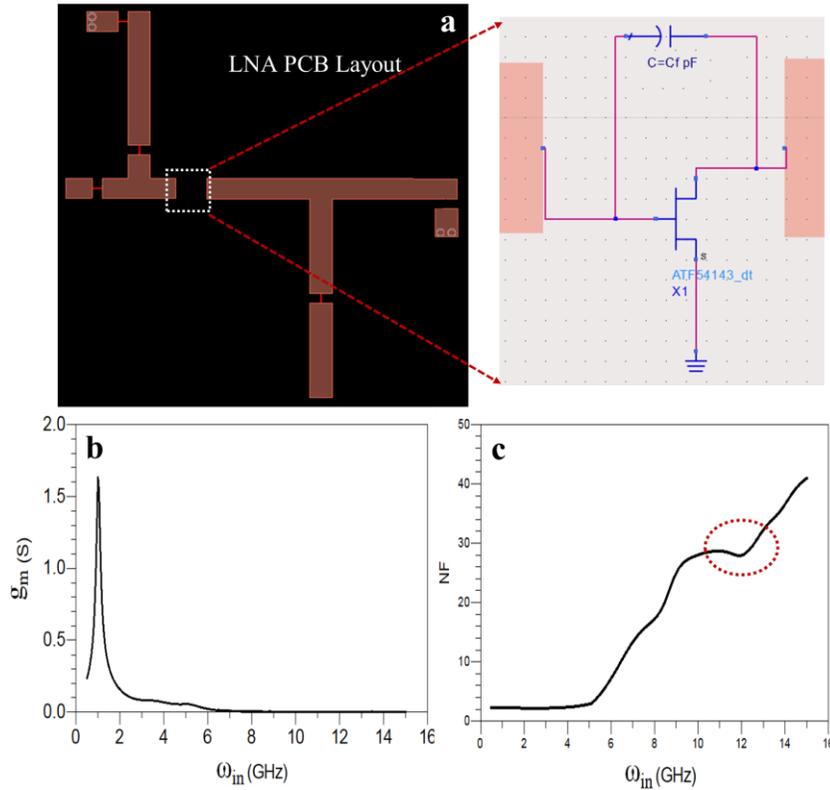

Fig. 5. Simulation of a LNA using ADS; a) PCB layout of the related LNA, b) LNA circuit transconductance vs. $\omega_{in}$ (GHz), c) noise figure of the LNA vs. $\omega_{in}$ (GHz)

The important result is illustrated in Fig. 5c where the NF is depicted versus incident wave frequency. In this figure as indicated with the dashed-ellipsoid on the figure, around 12 GHz the graph shows the odd behavior (a local minimum as a turning point). However, this figure doesn't show the NF changing the same as the Fig. 4. This contributes to the difference between the classical and quantum theory analysis. Nonetheless, the graph illustrated in Fig. 5c, shows a local minimum around the frequency that the summation of the two oscillator's frequencies equal to the incident frequency. This point partially confirms the odd behavior shown in Fig. 4. The mentioned deficiency is attributed to the transistor nonlinear model utilized for simulation of the LNA; in that model there is no degrees of freedom to change the model specifications such as transistor channel width and length, and also $SiO_2$ thickness. These are the critical parameters that we used from table 1 to model the LNA in the quantum realm.

Consequently, the idea of the oscillators photon number and its essential role to define the important characteristics of the LNA works well.

## IV. Conclusions

In this article, a LNA was designed and studied its operations using quantum theory. One of the main goals was that using quantum theory to analyze the LNA circuit gave us more degrees of freedom to efficiently manage the trade-off between quantities in the circuit. For this reason, we analyzed the circuit and derived the Lagrangian for the designed LNA and examined the Hamiltonian of the circuit. The Hamiltonian of the circuit was divided into two linear and nonlinear parts to completely clarify the circuit. The relationship showed that the designed LNA operated like two simple LC oscillators coupled to each other. Using this point, the energy levels of the two oscillators were calculated by applying the first order perturbation theory. Additionally, the effect of the perturbation Hamiltonian was investigated on the oscillator states. Finally, the voltage and current signals as the observable operators were calculated and they were completely expressed in terms of the oscillator's photon number. Using the derived relationship, a general formula was introduced for NF by which one can easily find the effect of the oscillator's number of photons on the NF and how the first and the second oscillators can affect NF. It was the main goal of this study that we were looking for. Additionally, the considered circuit was designed and classically simulated; the results partially testified the derived results using the quantum theory. Of course, we thought that this is due to the lack of the classical approach to fully analysis the circuit.

**Acknowledgment**
This work is supported by Çankaya University in Turkey.

**References**
[1] B. R. Mahafza, "Radar Systems Analysis and Design Using MATLAB," Chapman & Hall/CR, 2000.
[2] M.L. Skolnik, "Introduction to Radar Systems," Third Edition, McGraw Hill, 2001.
[3] B. T. Venkatesh Murthy, I. Srinivasa Rao, "Highly linear dual capacitive feedback LNA for Lband atmospheric radars," *Journal of Electromagnetic Waves and applications*, vol. 30, pp. 1-15, 2016.
[4] M. P. van der Heijden, L. C. N. de Vreede, and J. N. Burghartz, "On the Design of Unilateral Dual-Loop Feedback Low-Noise Amplifiers with Simultaneous Noise, Impedance, and IIP3 Match," *IEEE J.Solid-State Circuits*, vol. 39, pp. 1727-1737, 2004.
[5] V. Aparin and L. E. Larson, "Modified Derivative Superposition Method for Linearizing FET Low-Noise Amplifiers," *IEEE Trans. Microw. Theory Techn*, vol. 53, pp. 571-581, 2005.
[6] S. Ganesan, E. Sánchez-Sinencio, and J. Silva-Martinez, "A Highly Linear Low-Noise Amplifier, *IEEE Trans. Microw. Theory Techn*," vol. 54, pp. 4079-4086, 2006.
[7] X. Fan, H. Zhang, and E. Sánchez-Sinencio, "A Noise Reduction and Linearity Improvement Technique for a Differential Cascode LNA," *IEEE J.Solid-State Circuits,* vol. 43, pp. 588-599, 2008.
[8] D. K. Sheffer and T. H. Lee, "A 1.5 V, 1.5 GHz CMOS low-noise amplifier," *IEEE J. Solid-State Circuits*, vol. 32, pp. 745–759, 1997.
[9] D. K. Sheffer and T. H. Lee, "Corrections to A 1.5 V, 1.5 GHz CMOS low-noise amplifier," *IEEE J. Solid-State Circuits*, vol. 40, pp.1397–1398, 2005.
[10] X. Li, S. Shekhar, and D. J. Allstot, "Gm-boosted common-gate LNA and differential Colpitts VCO/QVCO in 0.18-*u*m CMOS," *IEEE J.Solid-State Circuits*, vol. 40, pp. 2609–2619, 2005.
[11] W. Zhuo, X. Li, S. Shekhar, S. H. K. Embabi, J. Pineda de Gyvez, D. J. Allstot, and E. Sánchez-Sinencio, "A capacitor cross-coupled common-gate low noise amplifier," *IEEE Trans. Circuits Syst. II: Expr. Briefs*, vol. 52, pp. 875–879, 2005.
[12] Z. Hamaizia, N. Sengouga, M. Missous, M.C.E. Yagoub, "A 0.4 dB noise figure wideband low-noise amplifier using a novel InGaAs/InAlAs/InP device," *Materials Science in Semiconductor Processing*, vol. 14, pp. 89–93, 2011.
[13] M. Lanzagorta, "Quantum Radar," A Publication in the Morgan & Claypool Publishers series, 2012.
[14] A. Salmanogli, D. Gokcen, "Entanglement Sustainability Improvement Using Optoelectronic Converter in Quantum Radar (Interferometric Object-Sensing)," *IEEE Sensors Journal*, vol. 21, pp. 9054-9062, 2021.
[15] A Salmanogli, D Gokcen, HS Gecim, "Entanglement Sustainability in Quantum Radar," *IEEE J. Sel. Top. Quantum Electron,* vol. 26, pp. 1-11, 2020.
[16] B. Huttner, S. M. Barnett, "Quantization of the electromagnetic field in dielectrics," *Phys Rev A*, vol. *46*, pp. 4306-14, 1992.
[17] M. O. Scully, M. S. Zubairy, "Quantum Optics," Cambridge University Press, UK, 1997.

**Appendix A:**
To deriving the Hamiltonian, the relationship between $[Q_1,Q_2]$ and $[\partial\varphi_1/\partial t, \partial\varphi_2/\partial t]$ is expressed as:

$$\begin{pmatrix} Q_1 \\ Q_2 \end{pmatrix} = \begin{pmatrix} C_{in}+C_{gd}+C_{gs}+C_N & -C_{gd} \\ -C_{gd} & C_d+C_{gd} \end{pmatrix} \begin{pmatrix} \dot\varphi_1 \\ \dot\varphi_2 \end{pmatrix} + \begin{pmatrix} 0 & g_m \\ 0 & 0 \end{pmatrix} \begin{pmatrix} \varphi_1 \\ \varphi_2 \end{pmatrix} + \begin{pmatrix} -C_{in}V_{rf} \\ 0 \end{pmatrix} \quad (A1)$$

where $C_N = (2g_{m2}+3g_{m3L})\times\varphi_{2DC}$ considered as the nonlinear capacitor and $\varphi_{2DC}$ stands for the amount of $\varphi_2$ at the operational point. Using (A1), one can express $[\partial\varphi_1/\partial t, \partial\varphi_2/\partial t]$ in terms of $[Q_1,Q_2]$ and $[\varphi_1,\varphi_2]$ as:

$$\begin{pmatrix} \dot\varphi_1 \\ \dot\varphi_2 \end{pmatrix} = \begin{pmatrix} C_{11} & C_{12} \\ C_{21} & C_{22} \end{pmatrix} \begin{pmatrix} Q_1 \\ Q_2 \end{pmatrix} - \begin{pmatrix} C_{11} & C_{12} \\ C_{21} & C_{22} \end{pmatrix} \begin{pmatrix} 0 & g_m \\ 0 & 0 \end{pmatrix} \begin{pmatrix} \varphi_1 \\ \varphi_2 \end{pmatrix} - \begin{pmatrix} C_{11} & C_{12} \\ C_{21} & C_{22} \end{pmatrix} \begin{pmatrix} -C_{in}V_{rf} \\ 0 \end{pmatrix} \quad (A2)$$

where $C_{11}, C_{12}, C_{21},$ and $C_{22}$ can be expressed as:

$$C_{11} = \frac{C_d+C_{gd}}{C_{inv}}; C_{12} = C_{21} = \frac{C_{gd}}{C_{inv}}; C_{22} = \frac{C_{in}+C_N+C_d+C_{gd}}{C_{inv}}$$

$$C_{inv} = (C_{in}+C_N+C_d+C_{gd})\times(C_d+C_{gd}) - C_{gd}^2 \quad (A3)$$

Also to simplify the algebra, we used definitions as:

$$\frac{1}{2C_{q1}} = \frac{C_{11}^2(C_{in}+C_{gs}+C_{gd})}{2} + \frac{C_{21}^2(C_d+C_{gd})}{2} - C_{gd}C_{21}C_{11}$$

$$\frac{1}{2C_{q2}} = \frac{C_{12}^2(C_{in}+C_{gs}+C_{gd})}{2} + \frac{C_{22}^2(C_d+C_{gd})}{2} - C_{gd}C_{12}C_{22}$$

$$\frac{1}{2C_{q1q2}} = \frac{2C_{12}C_{11}(C_{in}+C_{gs}+C_{gd})}{2} + \frac{2C_{22}C_{21}(C_d+C_{gd})}{2} - C_{gd}(C_{11}C_{22}+C_{21}C_{12})$$

$$\frac{1}{2C_{p1}} = \frac{C_{11}^2(C_{in}+C_{gs}+C_{gd})}{2}; \frac{1}{2C_{p2}} = \frac{C_{22}^2(C_d+C_{gd})}{2}; \frac{1}{2C_{q1p1}} = \frac{2C_{11}^2(C_{in}+C_{gs}+C_{gd})}{2} + C_{gd}C_{21}C_{11} \quad (A4)$$

$$\frac{1}{2C_{q2p2}} = -\frac{2C_{22}C_{21}(C_d+C_{gd})}{2} + C_{gd}C_{22}C_{21}; \frac{1}{2C_{q1p2}} = -\frac{2C_{21}^2(C_d+C_{gd})}{2} + C_{gd}C_{11}C_{12}$$

$$\frac{1}{2C_{p1p2}} = -C_{gd}C_{11}C_{12}; \frac{1}{2C_{q2p1}} = \frac{2C_{11}C_{12}(C_{in}+C_{gs}+C_{gd})}{2} + C_{gd}C_{22}C_{11}$$

$$\frac{1}{2C'_{p1p2}} = -2C_{11}^2C_{in}; \frac{1}{2C'_{q1p2}} = -2C_{11}^2C_{in}; \frac{1}{2C'_{q2p2}} = -2C_{11}C_{12}C_{in}$$

The other parameters are expressed as:

$$\frac{P_{11}}{2} = -2C_{11}^2C_{in}.\frac{(C_{in}+C_{gs}+C_{gd})}{2}; \frac{P_{12}}{2} = C_{gd}C_{21}C_{11}C_{in}; \frac{P_{21}}{2} = -2C_{21}^2C_{in}.\frac{(C_d+C_{gd})}{2};$$

$$\frac{P_{22}}{2} = C_{gd}C_{21}C_{11}C_{in}; \frac{q_{11}}{2} = -2C_{11}^2C_{in}.\frac{(C_{in}+C_{gs}+C_{gd})}{2}; \frac{q_{21}}{2} = -2C_{11}C_{12}C_{in}.\frac{(C_{in}+C_{gs}+C_{gd})}{2} \quad (A5)$$

$$\frac{q_{12}}{2} = 2C_{21}^2C_{in}.\frac{(C_d+C_{gd})}{2}; \frac{q_{22}}{2} = 2C_{21}C_{22}C_{in}.\frac{(C_d+C_{gd})}{2}; \frac{q_{13}}{2} = -C_{gd}C_{21}C_{11}C_{in}; \frac{q_{23}}{2} = -C_{gd}C_{22}C_{11}C_{in}$$

$$P_1 = P_{11}+P_{12} - \frac{\overline{I_s}}{g_mV_{rf}}; P_2 = P_{21}+P_{22} - \frac{\overline{I_d}}{g_mV_{rf}}; G_1 = q_{11}+q_{12}+q_{13}; G_2 = q_{21}+q_{22}+q_{23}$$

The constants $A_1, A_2, A_3, B_1, B_2, B_3, E_{1\omega},$ and $E_{2\omega}$ are defined as:

$$A_1 = \frac{1}{\omega_2}\left\{\frac{-i}{4C_{q1q2}\sqrt{Z_1Z_2}} - \frac{g_m}{4C_{q2p1}}\right\}; A_3 = \frac{1}{\omega_1}\left\{\frac{g_m}{4C_{q1p1}}\right\}$$

$$A_2 = \frac{1}{\omega_2}\left\{\frac{-ig_m^2}{4C_{p1p2}}\sqrt{Z_1Z_2} + \frac{g_m}{4C_{q1p2}}\sqrt{\frac{Z_2}{Z_1}} - \frac{ig_m g_{NL}V_{rf}}{4C'_{p1p2}}\sqrt{Z_1Z_2} + \frac{g_{NL}V_{rf}}{4C'_{q1p2}}\sqrt{\frac{Z_2}{Z_1}}\right\}$$

$$B_1 = \frac{1}{\omega_1}\left\{\frac{-i}{4C_{q1q2}\sqrt{Z_1Z_2}} - \frac{g_m}{4C_{q1p2}}\sqrt{\frac{Z_2}{Z_1}} - \frac{g_{NL}V_{rf}}{4C'_{q1p2}}\sqrt{\frac{Z_2}{Z_1}}\right\}; B_3 = \frac{1}{\omega_2}\left\{\frac{g_m}{4C_{q2p2}} + \frac{g_{NL}V_{rf}}{4C'_{q2p2}}\right\} \quad (A6)$$

$$B_2 = \frac{1}{\omega_1}\left\{\frac{-ig_m^2}{4C_{p1p2}}\sqrt{Z_1Z_2} + \frac{g_m}{4C_{q2p1}} - \frac{ig_m g_{NL}V_{rf}}{4C'_{p1p2}}\sqrt{Z_1Z_2}\right\}$$

$$E_{1\omega} = \frac{G_1 V_{rf}}{2}\sqrt{\frac{1}{2Z_1\hbar}} - \frac{iP_1 g_m V_{rf}}{2}\sqrt{\frac{Z_1}{2\hbar}}$$

$$E_{2\omega} = \frac{G_2 V_{rf}}{2}\sqrt{\frac{1}{2Z_2\hbar}} - \frac{iP_2 g_m V_{rf}}{2}\sqrt{\frac{Z_2}{2\hbar}} - iC_{11}^2 C_{in}^2 g_{NL} V_{rf}^2 \sqrt{\frac{Z_2}{2\hbar}}$$

where $g_{NL} = g_{m2} + 2g_{m3L}$.

**Appendix B:**
The S-parameters of the simulated LNA are presented as:

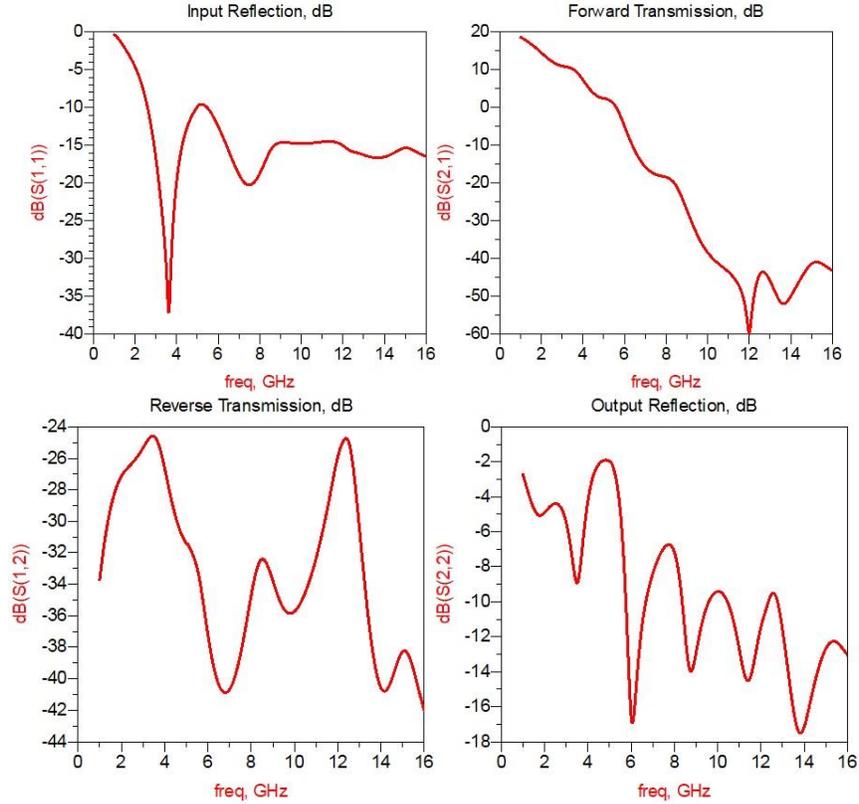